\documentclass[10pt,preprint,5p,twocolumn,times]{elsarticle}

\usepackage{amsmath,amssymb}
\usepackage{dcolumn} 
\usepackage{graphicx}
\usepackage{hyperref}
\usepackage{color}
\usepackage{colortbl}
\usepackage[tabel]{xcolor}
\usepackage{braket}
\usepackage{booktabs}
\usepackage{subfigure}
\usepackage{verbatim}
\usepackage{algorithmic}
\usepackage{algorithm}
\newcommand\BibTeX{{\rmfamily B\kern-.05em \textsc{i\kern-.025em b}\kern-.08em
T\kern-.1667em\lower.7ex\hbox{E}\kern-.125emX}}

\usepackage{multirow}
\usepackage{amssymb}
\usepackage{amsmath}
\usepackage{graphicx}
\usepackage{algorithmic}
\usepackage{algorithm}
\usepackage{color,graphicx,bm,multirow}
\renewcommand{\t}[1]{\textrm{\scriptsize #1}}

\begin{document}
\journal{Computer Physics Communications}
\begin{frontmatter}
\bibliographystyle{model1a-num-names}
\biboptions{square,numbers,comma,sort&compress}

\title{The dynamic parallel distribution algorithm for hybrid density-functional calculations in HONPAS package}

\author[CAS]{Honghui Shang\footnotemark[\value{footnote}]}
\author[CAS]{Lei Xu\footnotemark[\value{footnote}]}
\author[CAS]{Baodong Wu}
\author[USTC]{Xinming Qin}
\author[CAS]{Yunquan Zhang}
\author[USTC]{Jinlong Yang}

\address[CAS]{State Key Laboratory of Computer Architecture, Institute of Computing Technology, Chinese Academy of Sciences, Beijing}
\address[USTC]{Hefei National Laboratory for Physical
Sciences at Microscale, Department of Chemical Physics, and
Synergetic Innovation Center of Quantum Information and Quantum
Physics, University of Science and Technology of China, Hefei, Anhui
230026, China}

\begin{abstract}
This work presents a dynamic parallel distribution scheme for the Hartree-Fock exchange~(HFX) calculations based on the real-space NAO2GTO framework. The most time-consuming electron repulsion integrals~(ERIs) calculation is perfectly load-balanced with 2-level master-worker dynamic parallel scheme, the density matrix and the HFX matrix are both stored in the sparse format, the network communication time is minimized via only communicating the index of the batched ERIs and the final sparse matrix form of the HFX matrix. The performance of this dynamic scalable distributed algorithm has been demonstrated by several examples of large scale hybrid density-functional calculations on Tianhe-2 supercomputers, 
including both molecular and solid states systems with multiple dimensions, and illustrates good scalability.
\end{abstract}

\end{frontmatter}
\footnotetext{Both authors contributed equally to this work.}
\section{Introduction}
The hybrid density-functional calculations \cite{Becke_1993,Stephens_1994,Janesko_2009,Paier_2009,HSE-better-1,HSE-better-2,HSE-better-3,HSE03-1,HSE03-2,HSE06,Gaussian,VASP-HSE}, which contains the Hartree-Fock exchange~(HFX), shows the great accuracy performance for 
the geometry parameters, band structure properties and cohesive energies of 
a large range of materials. However, the computational time is extremely expensive compared to the conventional ground state density-functionals theory~(DFT) calculation duo to the calculation of the electron repulsion integrals (ERIs), which is the most time-consuming part in the HFX matrix construction. Therefore, a highly efficient and scalable implementations of the ERIs is urgently needed.

There have been a variety of implementation of the hybrid density-functionals for solid state physics calculations. We broadly classify
these works in the two categories by the usage of the basis set: plane waves~(PW) method\cite{Gonze2016,VASP-HSE,Lin2016,Gygi2010,DiStasio2014, DeSlippe2017, Natan2015,Natan2016,Wu2009} or linear combination of atomic orbitals~(LCAO) method\cite{HSE06,Gaussian,CRYSTAL}. The plane wave basis set is
the complete basis set, but not localized. On the contrary, the atomic orbitals basis sets are localized, which make the Hamiltonian matrices to be sparse. As a result, the atomic basis sets have attracted considerable interest for DFT calculations because of their favorable scaling with respect to the  number of atoms and their potential for massively parallel implementations for large-scale calculations~\cite{DMOL,SIESTA,Blum2009,Havu2009,Ren2012,GAPW,Mohr2014,Gaussian,Shang-JCP}. There are mainly two types of the atomic orbitals, 
one is the gaussian type orbital~(GTO), as adopted in Gaussian\cite{Gaussian} and CRYSTAL\cite{CRYSTAL}et al.; the other one is the numerical atomic orbital~(NAO), which is adopted in SIESTA\cite{SIESTA}, DMOL\cite{DMOL}, OPENMX\cite{OPENMX},FHI-aims\cite{Blum2009} et al.. The advantage of GTO is the analytical calculation of the ERIs, and
the advantage of NAO is its strict locality, which naturally leads to lower order scaling of computational time versus system size. We 
have proposed a mixed scheme called NAO2GTO\cite{Shang-JCP} to take advantages of both types of atomic orbitals. In the NAO2GTO method, 
the strict cutoff of the atomic orbitals is satisfied with NAO, and then the NAO is fitted with several GTOs to analytically calculate the ERIs, 
after employing several ERI screening techniques, the construction of HFX matrix can be very efficient and scale linearly\cite{Shang-JCP,Qin2015}.

In the parallelization of HFX matrix construction, we have to pay attention to two problems, one is the load balancing of the ERIs, and the other one is the communications of the density and HFX matrices.
Previously, the load balancing of ERIs were solved by static or dynamic distribution schemes\cite{Shang2019,NWChem,GTFock_1,GTFock_2,CRYSTAL,Quickstep,Weber2006}.

The major difference between static\cite{Shang2019,CRYSTAL,Weber2006} and dynamic\cite{NWChem,GTFock_1,GTFock_2,Quickstep} parallel
distribution algorithm is the way how to parallelize the computation of the ERIs. In the static parallel distribution algorithm, the ERIs are distributed among the processors before all the calculations of the ERIs; In the dynamic parallel distribution algorithm, the distribution and the calculation of the ERIs are performed simultaneously, which improves the load balance and parallel efficiency. For instance, the NWchem\cite{NWChem} software uses a simple centralized dynamic scheduling algorithm to distribute the ERIs to the worker processes, but the parallel efficiency decreases when very large numbers of processes are used. The GTFock\cite{GTFock_1,GTFock_2} code uses an initial static task partitioning scheme along with a work-stealing distributed dynamic scheduling algorithm, and it gives very good
parallel scalability for the ERIs calculations. In CP2K/Quickstep\cite{Quickstep}, the ERIs are coarse grained using bins, and then based on the estimated cost of each bin, the simulated annealing method is adopted to redistribute all the bins to improve the load balance, which is limited by the accuracy of the estimated cost of each bin.
In order to reduce the communication time, both NWchem and GTFock use the Global Arrays framework which provides the one-sided communication scheme to achieve high parallel performance with the distributed HFX matrix computations. On the contrary, the CP2K/Quickstep\cite{Quickstep} replicates the global density and HFX matrix on each MPI process in order to minimize the communication, however, it limits system size because as the memory usage scaling as O(N$^2$), and it also limits the parallel scalability as the synchronization of the HFX matrix becomes the bottleneck when using very large numbers of cores.

Recently, we have proposed two static distribution strategies\cite{Shang2019} for the calculation of the ERIs, however, the static distribution of ERI shell pairs produces load imbalance that causes the decreasing of the  parallel efficiency, while the static distribution of ERI shell 
quartet can yield very high load balance, but because of the need of the global ERI screening calculation, the parallel efficiency has also been dramatically reduced, that both of the static distribution schemes limiting parallel scalability.
In order to improve the parallel efficiency, the dynamic parallel distribution algorithm is needed.

Here in this work, a new dynamic parallel distribution algorithm based on the NAO2GTO scheme\cite{Shang-JCP} has been proposed and implemented in the Order-N performance HONPAS code\cite{Qin2015}.
In our approaches, the calculations of the ERIs are perfectly loading balanced and can scale to very large numbers of cores thanks to the 2-level master-worker distribution of shell pairs. Furthermore, the communication time is minimized by using the 
global sparse matrix with linear scaling memory usage.
The efficiency and scalability of these algorithms are demonstrated by benchmark timings in the periodic solid system with hundreds to thousands of atoms in the unit cell.

The remainder of the paper is organized as follows. In Sec.~\ref{sec:background} we succinctly summarize the fundamental background of this study. Then the dynamic parallel scheme and the detailed implementation of our parallel distribution strategies are discussed in Sec.~\ref{sec:parallel}. Furthermore, we demonstrated the parallel scalability of our implementation in Sec.~\ref{sec:result}. Finally, Sec.~\ref{sec:conclusion} summarizes the main ideas and findings of this work.

\section{Background}
\label{sec:background}
In this section, we recall the basis theoretical framework used in this work.  
A spin-unpolarized notation is used throughout the text for the sake of simplicity, but a formal generalization to the collinear spin case is straightforward. The total-energy in the Kohn-Sham DFT is defined as 
\begin{equation}
E_\t{KS}= T_\t{s}[n]+E_\t{ext}[n]+E_\t{H}[n]+E_\t{xc}[n] + E_\t{nuc-nuc}\;.
\label{eq:KSTOT}
\end{equation}
Here, $n(\mathbf{r})$ is the electron density, $T_\t{s}$ is the kinetic energy of non-interacting electrons, while $E_\t{ext}$ is external energy stemming from the electron-nuclear attraction, $E_\t{H}$ is the Hartree energy, $E_\t{xc}$ is the exchange-correlation energy, and $E_\t{nuc-nuc}$ is the nucleus-nucleus repulsion energy. The ground state electron density~$n_0(\mathbf{r})$ (and the associated ground state total energy) 
is obtained by variationally minimizing Eq.~(\ref{eq:KSTOT}) under the constraint that
the number of electrons $N_e$ is conserved. This yields the chemical potential $\mu=\delta E_{KS}/\delta n$ of the electrons and the Kohn-Sham single particle equations
\begin{equation}
\hat{h}_\t{KS}\psi_i = \left[ \hat{t}_\t{s} + v_\t{ext}(r)+v_\t{H}+v_\t{xc}\right] \psi_i = \epsilon_{p} \psi_i \;,
\label{eq:ks-equation}
\end{equation}
for the Kohn-Sham Hamiltonian~$\hat{h}_\t{KS}$. In Eq.~(\ref{eq:ks-equation}), $\hat{t}_\t{s}$ denotes the kinetic energy operator, $v_\t{ext}$ the external potential, $v_{H}$ the Hartree potential, and $v_{xc}$ the exchange-correlation potential. Solving Eq.~(\ref{eq:ks-equation}) yields the Kohn-Sham single particle states~$\psi_p$ and their eigenenergies~$\epsilon_{p}$.
The single particle states determine the electron density via
\begin{equation}
n(\mathbf{r})=\sum_i f_i |\psi_i|^2 \;,
\end{equation}
in which $f_i$ denotes the Fermi-Dirac distribution function,  and $i$ is the suffix for different Kohn-Sham state.

The Eq.~(\ref{eq:ks-equation}) can be solved numerically by expanding the Kohn-Sham states $\psi_{i}$ with a finite basis set. In periodic systems, such Kohn-Sham states are also called crystalline orbitals, which are normalized in the full space with a linear combination of Bloch functions
$\phi_{\mu}(\mathbf{k,r})$ to satisfy the periodic boundary condition,
\begin{equation}
\psi_{i}(\mathbf{k,r})=
\sum_{\mu}c_{\mu,i}(\mathbf{k})\phi_{\mu}(\mathbf{k,r}) \;.
\end{equation}
Such Bloch functions are defined in terms of atomic orbitals $\chi_{\mu}^{\mathbf{R}}(\mathbf{r})$.\\
\begin{equation}
\phi_{\mu}(\mathbf{k,r})=\dfrac{1}{\sqrt{N}}\sum_{\mathbf{R}}\chi_{\mu}^{\mathbf{R}}(\mathbf{r})\mathbf{e}^{i\mathbf{k}\cdotp
(\mathbf{R+r_{\mu}})} \;,
\end{equation}
where the Greek letter $\mu$ is the index of atomic orbitals, $\mathbf{R}$ is the origin of the auxiliary supercell, N is the number of unit cells in the system.
$\chi_{\mu}^{\mathbf{R}}(\mathbf{r})=\chi_{\mu}(\mathbf{r-R-r_\mu})$
is the $\mu$-th atomic orbital, whose center is displaced from the origin of the auxiliary supercell  at $\mathbf{R}$ by $\mathbf{r}_\mu$. $c_{\mu,i}(\mathbf{k})$ is the wave function coefficient, which is
obtained by solving the following generalized eigenvalue equation,

\begin{equation}
H(\mathbf{k})c(\mathbf{k})=E(\mathbf{k})S(\mathbf{k})c(\mathbf{k})\;,
\end{equation}
where
\begin{equation}
 [H(\mathbf{k})]_{\mu\nu}=\sum_{\mathbf{R}}<\chi_{\mu}^{\mathbf{0}}|\hat{H}|\chi_{\nu}^{\mathbf{R}}>  \mathbf{e}^{i\mathbf{k}\cdotp (\mathbf{R+r_{\nu}-r_{\mu}})} \;,
\end{equation}
and 
\begin{equation}
 [S(\mathbf{k})]_{\mu\nu}=\sum_{\mathbf{R}} <\chi_{\mu}^{\mathbf{0}}|\chi_{\nu}^{\mathbf{R}}> \mathbf{e}^{i\mathbf{k}\cdotp (\mathbf{R+r_{\nu}-r_{\mu}})} \;.
\end{equation}

The Hamiltonian matrix can be distributed into two parts, one is the conventional DFT part called
$\rm {H^{DFT}}$, and the other is $\rm{H^{HFX}}$ part which contains the calculation of the ERIs  
\begin{equation}
  [H^{\rm DFT}]_{\mu\lambda}^{\mathbf{G}}=<\chi_{\mu}^{\mathbf{0}}|  \hat{t}_\t{s} + v_\t{ext}(r)+v_\t{H}+v_\t{xc}   |\chi_{\lambda}^{\mathbf{G}}> \;,
  \label{eq:H-DFT}
\end{equation}
\begin{equation}
  [H^{\rm HFX}]_{\mu\lambda}^{\mathbf{G}}=-\frac{1}{2}\sum_{\nu\sigma}\sum_{\mathbf{N,H}}P_{\nu\sigma}^\mathbf{H-N}\mathbf{[(\chi_{\mu}^{0}\chi_{\nu}^{N}|\chi_{\lambda}^{G}\chi_{\sigma}^{H})]} \;,
    \label{eq:H-HFX}
\end{equation}
where $\mathbf{G}$, $\mathbf{N}$, and $\mathbf{H}$ represent the different origin of the auxiliary supercell~(the unit cell indexes), and the Greek letters $\mu,\lambda,\nu,\sigma$ represent the indexes of atomic orbitals.  Here the $P_{\nu\sigma}^\mathbf{N}$ denotes the density matrix which is computed by an integration of the wave function coefficient over the Brillouin zone (BZ) using 
\begin{equation}
 P_{\nu\sigma}^{\mathbf{N}}=\sum_j \int_{BZ} c_{\nu,j}^{*}(\mathbf{k})c_{\sigma,j}(\mathbf{k}) \theta(\epsilon_F-\epsilon_j(\mathbf{k}))\mathbf{e}^{i\mathbf{k}\cdotp \mathbf{N}}d\mathbf{k} \;,
\label{eq:dm}
\end{equation}
where $\theta$ is the step function, $\epsilon_F$ is the fermi
energy and $\epsilon_j(\mathbf{k})$ is the $j$-th eigenvalue at
point $\mathbf{k}$. 
And the full-range ERI is defined as   
\begin{equation}
 \mathbf{(\chi_{\mu}^{0}\chi_{\nu}^{N}|\chi_{\lambda}^{G}\chi_{\sigma}^{H})=\int\int   \frac{\chi_{\mu}^{0}(r)\chi_{\nu}^{N}(r)\chi_{\lambda}^{G}(r')\chi_{\sigma}^{H}(r')}{|r-r'|}
 drdr'} \;.
\end{equation}
For screened hybrid functional calculation, such as HSE06, only the short range part of the ERIs is needed,
\begin{equation}
 E_{\rm xc}^{\rm HSE06}=\dfrac{1}{4}E_{x}^{\rm SR-HF}(\omega)+\dfrac{3}{4}E_{x}^{\rm SR-PBE}(\omega)
+E_{x}^{\rm LR-PBE}(\omega)+E_{c}^{\rm PBE} \;,
\end{equation}
where  $\omega$=$0.11 Bohr^{-1}$ and $\mathbf{erfc}(r)=\dfrac{2}{\sqrt{\pi}}\int_r^\infty{e^{-t^2}dt}$.
The short-range and long-range part of PBE exchange functional is calculated following the Ref.\cite{HSE06}. 
All the ERIs' calculation of the following paper is for the short-range part,~i.e.
\[
  \mathbf{\mathbf{(\chi_{\mu}^{0}\chi_{\nu}^{N}|\chi_{\lambda}^{G}\chi_{\sigma}^{H})}_{\rm SR}} =  
  \]
\begin{equation}
   \mathbf{ \int\int \frac{\chi_{\mu}^{0}(r)\chi_{\nu}^{N}(r)erfc(\omega |r-r'|) \chi_{\lambda}^{G}(r')\chi_{\sigma}^{H}(r')}{|r-r'|}
 drdr'}   \;. 
 \label{eq:ERI-SR}
\end{equation}

It should be noted that this work focuses on the short-range HFX because the current auxiliary supercell is typically determined by the extent of the numerical orbitals, which is only valid for the short-range HFX. However, for the full HFX, this may not be enough since a larger auxiliary supercell is required for convergence.

After building the whole Hamiltonian, the KS wave function coefficients $c_{\mu,i}(\mathbf{k})$ are calculated using the standard diagonalization scheme, and finally we have the density matrix by using Eq.\ref{eq:dm}.

\begin{figure}
  \includegraphics[width=0.4\textwidth]{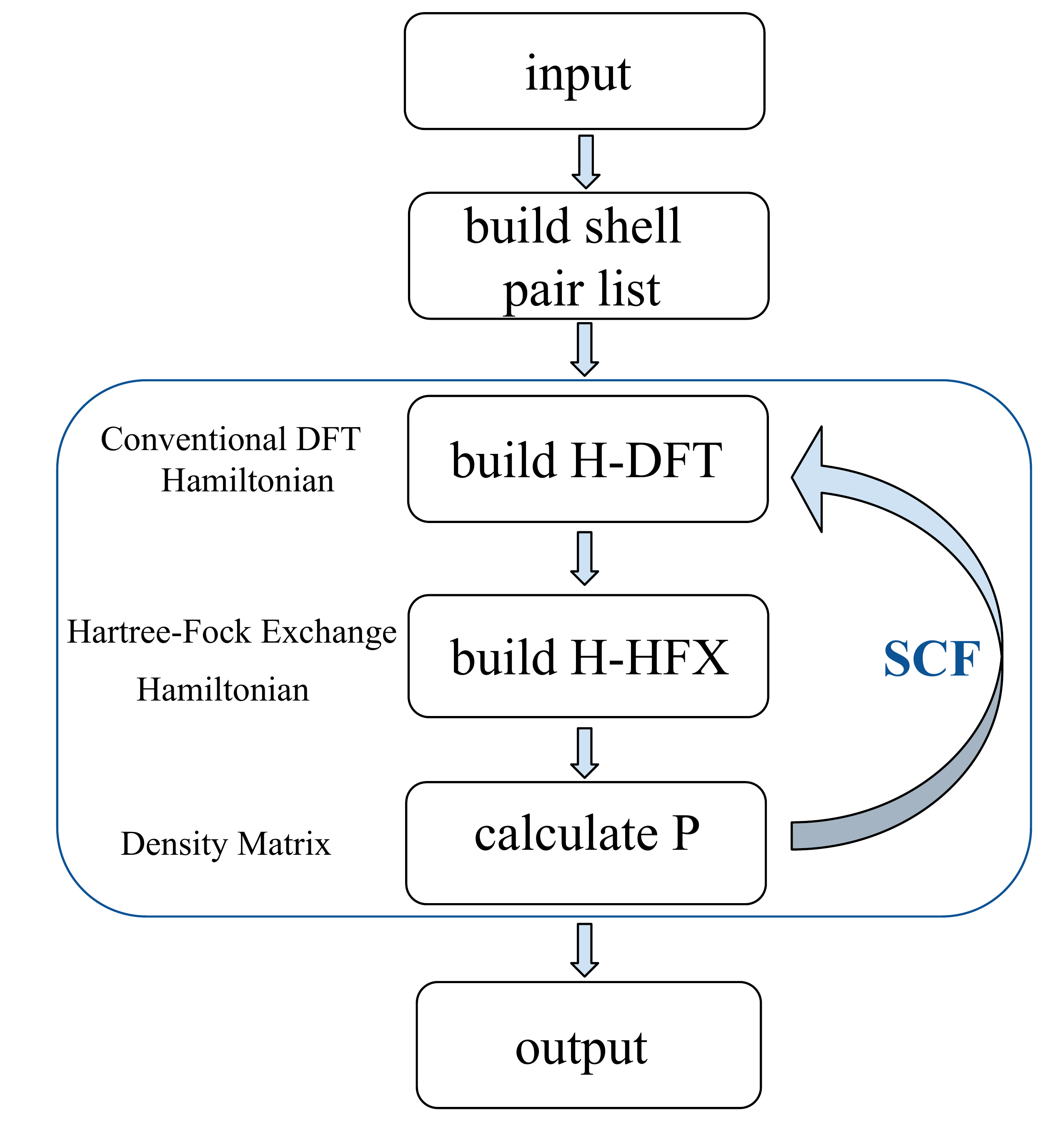}
  \caption{The flowchart of the hybrid functional calculation in the linear combination of atomic orbitals~(LCAO) approach.}  
  \label{fig:flowchart}
\end{figure}

The above procedures are repeated until the change of the density matrix element is smaller than a threshold, then we get a converged density and Hamiltonian matrices in the hybrid functional calculation as shown in Fig.\ref{fig:flowchart}.

In order to make the calculation of the ERIs more efficient, we have adopted the following computational schemes. 
Firstly, in our implementation, the 8-fold full permutation
symmetry of the ERIs has been considered for both the molecules and the solids systems, 
and in this way, we have a speedup of a factor 8 for the CPU time and a memory reduction of the same size.
\[
 (\mu^{\mathbf{0}}\nu^{\mathbf{H}}|\lambda^{\mathbf{G}}\sigma^{\mathbf{N}})=
   (\mu^{\mathbf{0}}\nu^{\mathbf{H}}|\sigma^{\mathbf{N}}\lambda^{\mathbf{G}})=
\]
\[
   (\nu^{\mathbf{0}}\mu^{\mathbf{-H}}|\lambda^{\mathbf{G-H}}\sigma^{\mathbf{N-H}})=
   (\nu^{\mathbf{0}}\mu^{\mathbf{-H}}|\sigma^{\mathbf{N-H}}\lambda^{\mathbf{G-H}})=
\]
\[
   (\lambda^{\mathbf{0}}\sigma^{\mathbf{N-G}}|\mu^{\mathbf{-G}}\nu^{\mathbf{H-G}})=
   (\lambda^{\mathbf{0}}\sigma^{\mathbf{N-G}}|\nu^{\mathbf{H-G}}\mu^{\mathbf{-G}})=
\]
\begin{equation}
   (\sigma^{\mathbf{0}}\lambda^{\mathbf{G-N}}|\mu^{\mathbf{-N}}\nu^{\mathbf{H-N}})=
   (\sigma^{\mathbf{0}}\lambda^{\mathbf{G-N}}|\nu^{\mathbf{H-N}}\mu^{\mathbf{-N}})\;.
\end{equation}

Secondly, our NAO2GTO scheme\cite{Shang-JCP} is adopted to calculate the ERIs analytically with fitted GTOs, and as the angular part of the NAOs is spherical harmonic while the GTOs are Cartesian harmonic function, a transformation\cite{Schlegel} between the Cartesian and spherical harmonic functions are performed. After the transformation, the GTOs are grouped into shells according to the NAOs' angular momentum, thus, if $\mu \in I$, $\nu \in J$, $\lambda \in K$,  $\sigma \in L$, for the I, J, K, L shell quartet, then
all the integrals $(\mu\nu|\lambda\sigma)$ are computed together for one shell quartet at a time.
As a result, the computational expense is strongly dependent on the angular momenta of the shell quartet which needs to be distributed in parallel.

Thirdly, before the SCF cycle, two shell pair lists (list-IJ and list-KL) are firstly preselected according to Schwarz screening\cite{Schwarz}, as shown in Fig.\ref{fig:flowchart}
\begin{equation}
|(\mu\nu|\lambda\sigma)| \leqslant \sqrt{(\mu\nu|\mu\nu)(\lambda\sigma|\lambda\sigma)} \;,
\label{eq:schwarz}
\end{equation}
and only the shell list indexes with $(IJ|IJ)>\tau$ or $(KL|KL)>\tau$ (here $\tau$ is the drop tolerance) are stored.
As shown in Eq. (\ref{eq:H-HFX}), the first index I runs only within the unit cell, while the indexes (J,K,L) run over the whole supercell, so the list-IJ is smaller than the list-KL. Then in the ERIs calculations, the loops run over these two shell lists. 
Then before the calculation of every ERI, we use Schwarz inequality Eq. (\ref{eq:schwarz}) again to
estimate a rigorous upper bound, that only the ERIs with non-negligible contributions are calculated,
we note this screening method as Schwarz screening. Because the exponential decay of the charge distributions, the Schwarz screening reduces the total number of ERIs to be computed from $O(N^4)$ to $O(N^2)$. 
In addition to Schwarz screening, the NAO screening\cite{Shang-JCP} and the distance screening\cite{Shang-JCP} is also adopted to reduce the total number of ERIs from $O(N^2)$ to $O(N)$. 

Finally, we use the density matrix screening to further reduce
the number of ERIs, that the maximal value of the density matrix of each shell~($P_{max}$) is calculated during every SCF cycle, and then
the density matrix screening is,
\begin{equation}
P_{screening}\times \sqrt{(IJ|IJ)(KL|KL)}  \leqslant \epsilon_{Schwarz}\;,
\label{eq:dmscreening}
\end{equation}
where $P_{screening}=max({|P_{max}^{IK}|,|P_{max}^{IL} |,|P_{max}^{JK}|,|P_{max}^{JL}|})$ 
Here four density matrix elements are needed for the maximal value because of the 8-fold full permutation symmetry of the 
ERIs is exploited in the implementation. The maximal density matrix elements are chosen from the density matrix of the previous SCF cycle,
which produce a stable direct SCF cycle\cite{Almlof1982}.

\section{Parallelization strategies}
\label{sec:parallel}
In order to solve the contradiction between parallel efficiency and load imbalance in the static distribution strategy,
the dynamic load balancing scheme is adopted based on the master/worker method, that one of these processes is responsible 
for managing the distribution of all the ERIs , which is called  the master, as shown in Algorithm \ref{master process}, and the 
other processes compute the assigned ERIs, which are called the workers, as shown in Algorithm \ref{worker process}.

We can choose to assign only one task to the worker at a time, that the worker process requests only one ERI shell quartet from the 
master at one time and after receiving it, proceeds to compute it. However, such a scheme introduces too much communication time, and it will increase the execution time so as to reduces the parallel efficiency. As a result, here we choose to assign more than one ERI shell quartets at a time from the master process to the worker processes. Such a set of tasks is called the batched ERIs, and only the
start and end indexes of the batched ERIs are communicated. In practice, we use the receiver-initiated method. The task distribution procedure is initiated by the worker, which requests tasks from the master. Then the master chooses to send the indexes of the batched ERIs or terminal token based on whether there are tasks left or not. The worker who receives the task executes the task immediately and then requests the task after execution. If the worker receives a terminal token, it jumps out of the loop and ends the program. The master also exits the program after determining that the terminal token was sent to each worker. Such a master-worker scheme has been implemented using point-to-point blocking send and receive operations, and the full permutational symmetry of the ERIs has been considered.

\begin{algorithm}
  \caption{Flowchart of Master algorithms for ERIs.  N means the number of the ERIs in one batch, $N_{workers}$ means the number of the workers in the mater-worker scheme.}
  \begin{algorithmic} 
      \STATE MPI\_IRECV (to accept request) 
   \WHILE{$TaskCount > 0$}
   \IF{request detected}
   \STATE send the MESSAGE~(indexes of the batched ERIs)   
        \STATE $\rm TaskCount = TaskCount -N$ 
        \STATE MPI\_IRECV (to accept request) 
   \ENDIF       
   \ENDWHILE
    \IF{request detected}
   \STATE send terminal token
   \ENDIF
   \FOR{$i = 0$; $i < N_{workers} - 1$; $i ++$ }
   \STATE MPI\_IRECV (to accept request) 
   \IF{request detected}
   \STATE send terminal token
   \ENDIF
   \ENDFOR
  \label{master process}
  \end{algorithmic}
 \end{algorithm}

 \begin{algorithm}
	\caption{Flowchart of worker algorithms for ERIs.}
	\begin{algorithmic} 
		\WHILE{.TRUE.}
		\STATE send a task request 
		\STATE receive MESSAGE~(indexes of the batched ERIs) 
		\IF{MESSAGE is terminal token}
		\STATE exit
		\ENDIF
		\STATE compute the batched ERIs
		\ENDWHILE
	\end{algorithmic}
	\label{worker process}
\end{algorithm}

Although we can simply increase our computing power by increasing the number of workers, this increase is not infinite. Because the master process can only distribute one task at one time. When there are multiple task requests, a task request cannot be satisfied until the master has processed requests before it. This bottleneck will limit the efficiency of large-scale parallelism. In our test, the performance of single-level master-worker parallelism began to decreases when the 4000 cores were used, and the parallel efficiency of the ERIs calculation decreases to only 87\% when 10000 cores were used. This is because the master process is too busy to assign the tasks, resulting in the performance bottlenecks.

\begin{figure}
  \includegraphics[width=0.5\textwidth]{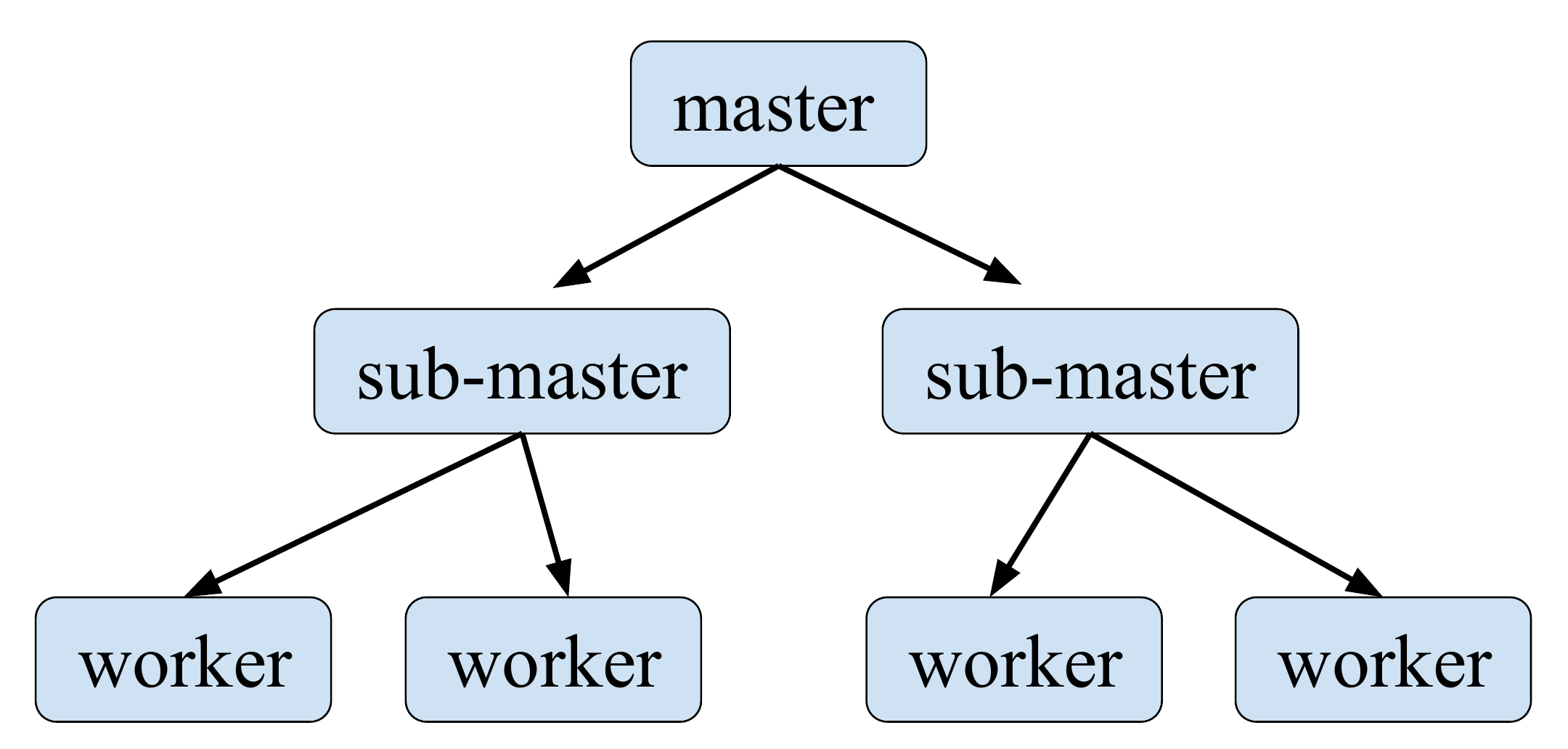}
  \caption{The illustration of the 2-level master-worker dynamic load balancing scheme.}
  \label{fig:multilevel}  
\end{figure}

In order to solve this problem, our approach is to distribute tasks with 2-level master-worker parallel algorithm, and we add a set of ``sub-master" processes between the master and workers as shown in Fig.\ref{fig:multilevel}. Each sub-master controls a group of workers and all sub-masters share the workload of master. Thus, multiple task fragments can be sent by multiple sub-masters at one time. When master and worker processes are almost unchanged, the sub-master process consists of three steps: request task, send task and close worker process, as shown in Algorithm \ref{sub-master process}. The parameters are more complex than the single-level distribution tasks mentioned above. In addition to considering how many tasks are sent between levels, we also need to consider how many sub-masters are set up and how many workers each sub-master has to manage. In our approach, each sub-master has been assigned to around 100 worker processes.

\begin{algorithm}
 \caption{Flowchart of sub-master algorithms for ERIs. N means the number of the ERIs in one batch, $N_{workers}$ means the number of the workers. }
 \begin{algorithmic}
  \WHILE{.TRUE.}
  \STATE send a task request to Master
  \STATE receive MESSAGE~(indexes of the batched ERIs or terminal token)
  \IF{MESSAGE is terminal token}
  \STATE exit
  \ENDIF
  \STATE MPI\_IRECV (to accept request)
  \WHILE{$TaskCount > 0$}
  \IF{request detected}
  \STATE send the MESSAGE~(indexes of the batched ERIs) 
  \STATE $\rm TaskCount = TaskCount -n$ 
  \STATE MPI\_IRECV (to accept request)
  \ENDIF      
  \ENDWHILE 
  \ENDWHILE
 \IF{request detected}
   \STATE send terminal token
   \ENDIF
   \FOR{$i = 0$; $i < N_{workers} - 1$; $i ++$ }
   \STATE MPI\_IRECV (to accept request) 
   \IF{request detected}
   \STATE send terminal token
   \ENDIF
   \ENDFOR
 \end{algorithmic}
 \label{sub-master process}
\end{algorithm}

In order to reduce the memory usage and minimize the communication time, the sparse format of both the density matrix and the HFX matrix
are replicated, which are much smaller than the dense matrix objects, as shown in Table 1, for instance, the dense format matrix of TiO$_2$ system has 19927296 elements which is 34 times larger than the sparse matrix format with 578616 elements. The sparse density matrices can be accessed by every MPI process, so after the worker processes get the indexes of the grouped ERIs that need to be calculated, the corresponding local HFX matrices are built using such global density matrices, and finally the {\rm MPI\_ALLREDUCE} operation is adopted to build the global HFX matrix. It should be noted that,
during the construction of the HFX matrix, the transformation between the sparse matrix index and the dense matrix index need to performed twice, one time for the read from the sparse density matrix,
the other one time for the write into the sparse HFX matrix.  The flowchart for the HFX matrix construction is shown in Algorithm~\ref{algo:globle-matrix-parallel}, which loops over shell pair lists.  

\begin{algorithm}
\caption{Flowchart of the HFX matrix construction. $I,J,K,L$ are for shell indexes.
$\rm{P}_{gs}$ is the global sparse density matrix, $\rm H^{HFX}_{gs}$ is the global sparse HFX matrix.  }
\begin{algorithmic}
 \FOR{shell list-$IJ$ and list-$KL$}
       \IF{shell ERI $(IJ|KL)$  is not screened }
           \STATE  compute shell ERI $(IJ|KL)$
           \STATE  transform dense matrix indexes to sparse matrix index  
           \STATE  get $\rm H^{HFX}_{gs}$ using  $(IJ|KL)$ and $\rm{P}_{gs}$
       \ENDIF 
 \ENDFOR
 \STATE  MPI\_AllReduce to get  $\rm H^{HFX}_{gs}$ 
\end{algorithmic}
\label{algo:globle-matrix-parallel}
\end{algorithm}

\section{Performance Results}
\label{sec:result}
All the results are calculation on the Tianhe-2 supercomputer located at the National Supercom-
puting Center in Guangzhou, China, which was developed by the National University of Defense Technology, China. Tianhe-2 is composed of 17920 nodes with a custom interconnect called TH Express-2 using a fat-tree topology. Each node is composed of two Intel Ivy Bridge E5-2692 processors (12 cores each at 2.2 GHz) and three Intel Xeon Phi 31S1P coprocessors (57 cores at 1.1 GHz). Memory on each node is 64 GB DRAM and 8 GB on each Intel Xeon Phi card. Capable of a peak performance of 54.9 PFlops, Tianhe-2 has achieved a sustained performance of 33.9 PFlops with a performance-per-watt of 1.9 GFlops/W. Tianhe-2 has 1.4 PB memory, 12.4 PB storage capacity, and power consumption of 17.8 MW. The larges number of nodes that we can use for performance test is 2150~(51600 cores), and only the Intel Xeon Ivy Bridge CPUs are adopted in this work. 
Since HONPAS is developed in the framework of SIESTA code,  only the norm-conserving pseudopotentials can be adopted. In the following calculations, the norm-conserving pseudopotentials generated with the Troullier-Martins\cite{TM} scheme, in fully separable form developed by Kleiman and Bylader\cite{KB}, are used to represent
interaction between core ion and valence electrons.  The screened hybrid
functional HSE06\cite{HSE06} was used in the all the calculations. The size of the batched ERIs is set to 2000000 in the master processor, and is set to 10000 in the sub-master processors.

The performance of our method is demonstrated using the instances of the DNA, titanium dioxide surface, and silicon solid to test the strong scaling of the HONPAS code, which is measured  by the change in CPU time with the number of core used to make the construction of the HFX matrix. The time measurements are for the HFX matrix construction in a single SCF step, including the time 
used to setup the $P_{max}$ for density matrix screening, to calculate the ERIs, and to sum up and redistribute the global sparse HFX matrix. It should be noted that, the $P_{max}$ time is a constant value, and takes very small fraction of the total time for these systems when the CPU cores are smaller than 1000. The time for synchronization of the HFX matrix is increase with the number of cores, and the  fraction  of this part is also increased. The sample of the test systems are listed in Tabel 1. These three examples have been chosen as they range from the one to three dimensional, and they are the typical applications in the materials science community.

\begin{table*}
\label{tab:systems}
\begin{tabular}{c | c c | c c | c}
\hline \hline
System &  Atoms in unit cell  &  Basis  in unit cell & Atoms in supercell &  Basis  in supercell  & Elements in sparse matrix \\
\hline
1D-DNA &  715                &     7183                 &715            & 7183      &  3500871         \\
2D-TiO$_2$ &      144         &     1488                & 1296               & 13392     &    578616   \\
3D-Si-SZ     &      2000         &    8000          &     2000        &    8000  & 2116000 \\
3D-Si-DZP    &      512         &        6656       &       512     &  6656     &  5016064 \\
\hline \hline
\end{tabular}
\caption{The systems used in this work.}
\end{table*}

\begin{figure}
  \includegraphics[width=0.5\textwidth]{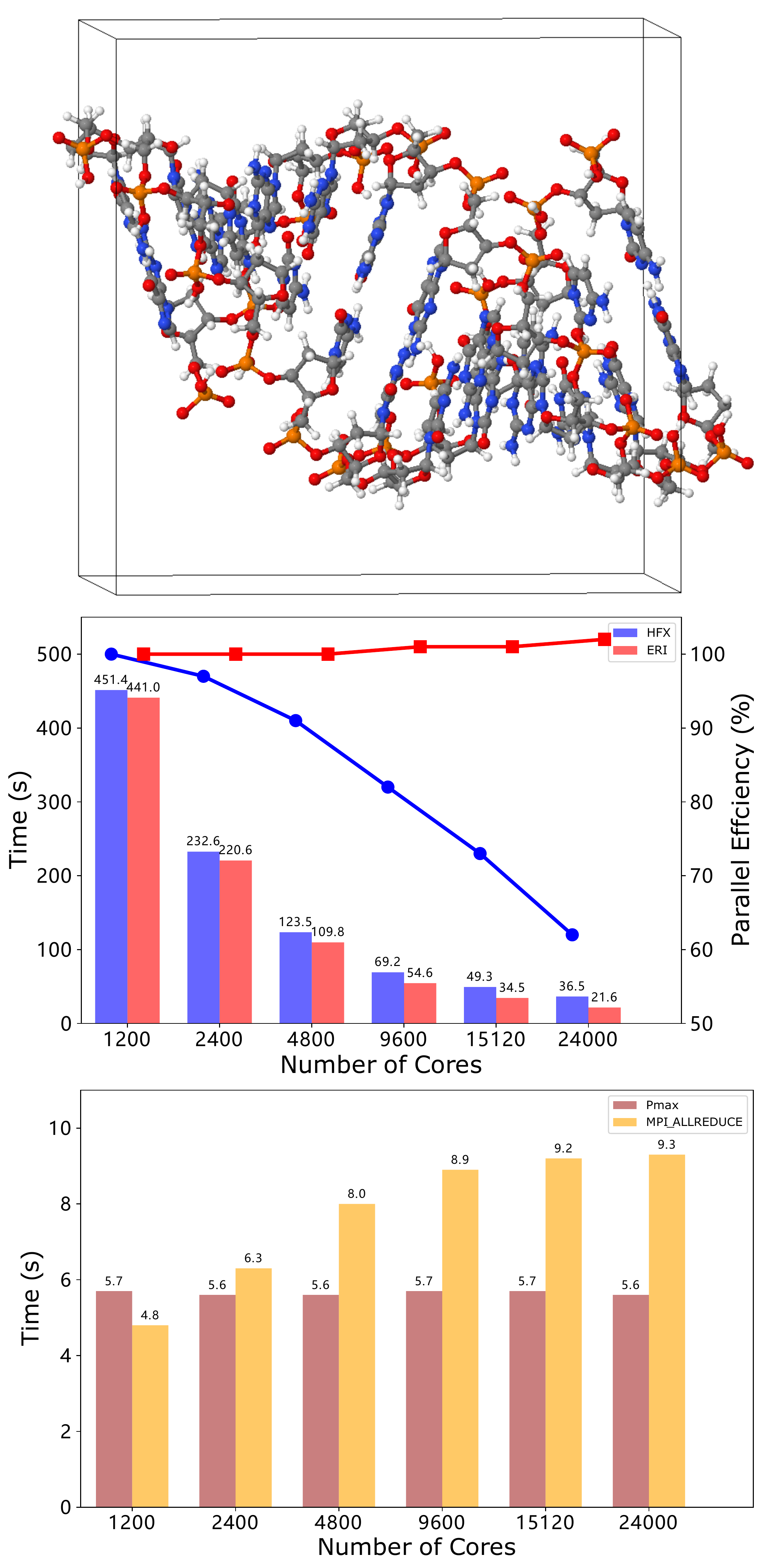}
    \caption{The strong scalability for the periodic DNA system. The blue/red/brown/orange bars correspond to the
    simulation time for the total HFX construction/ERI calculation/$P_{max}$ construction/MPI\_ALLREDUCE. The parallel efficiency of the HFX construction is labels with blue circles while the parallel efficiency of the ERIs calculations is labels with red squares. The time are annotated on top of the bars. The difference between the HFX time and the ERI time comes from the contributions from the $P_{max}$ selection process and the {\rm MPI\_ALLREDUCE} operation for the HFX matrix which are shown in the lower panel. }  
  \label{fig:DNA}
\end{figure}

The first system is the DNA contained 715 atom in the unitcell. The P, H, C, N and O  atoms are described using double-$\zeta$ plus polarization (DZP) valence basis sets yielding 7183 atomic orbitals per unitcell which is the rank of the Fock matrix. One k-point is used to sample the reciprocal space due to the large unit cell. In Fig.~\ref{fig:DNA}, the scalability of the HFX construction in one SCF cycle is presented and separated into
its three major components: calculation of the two-electron integrals (i.e. ERI), the calculation of the maximal values of density matrix in 
each shell at every SCF cycle(i.e. $P_{max}$) and the global summation of the HFX matrix(i.e. MPI\_ALLREDUCE). 
The scalability is good, especially almost ideal scaling is achieved for the calculation of the ERIs, which takes almost 98\% time with 1200 cores, on the other hand, the time for $P_{max}$ and  MPI\_ALLREDUCE do not scale with the number of CPU cores, so as the number of the CPU cores increased, the $P_{max}$ time percentage change from 1\% to 15\%, while the MPI\_ALLREDUCE time percentage change from 1\% to 25\%. Although the  $P_{max}$ and MPI\_ALLREDUCE are responsible for a small fraction of the total runtime, it is clear that the scaling of the  $P_{max}$ and MPI\_ALLREDUCE ultimately limits the final parallel scaling of the total HFX calculation.  As a result, despite the parallel efficiency of ERIs at 24000 cores is nearly 100\%, the parallel efficiency of the total HFX time which including ERI,  $P_{max}$, and MPI\_ALLREDUCE is only 61\% at 24000 cores.  
 
\begin{figure}
  \includegraphics[width=0.5\textwidth]{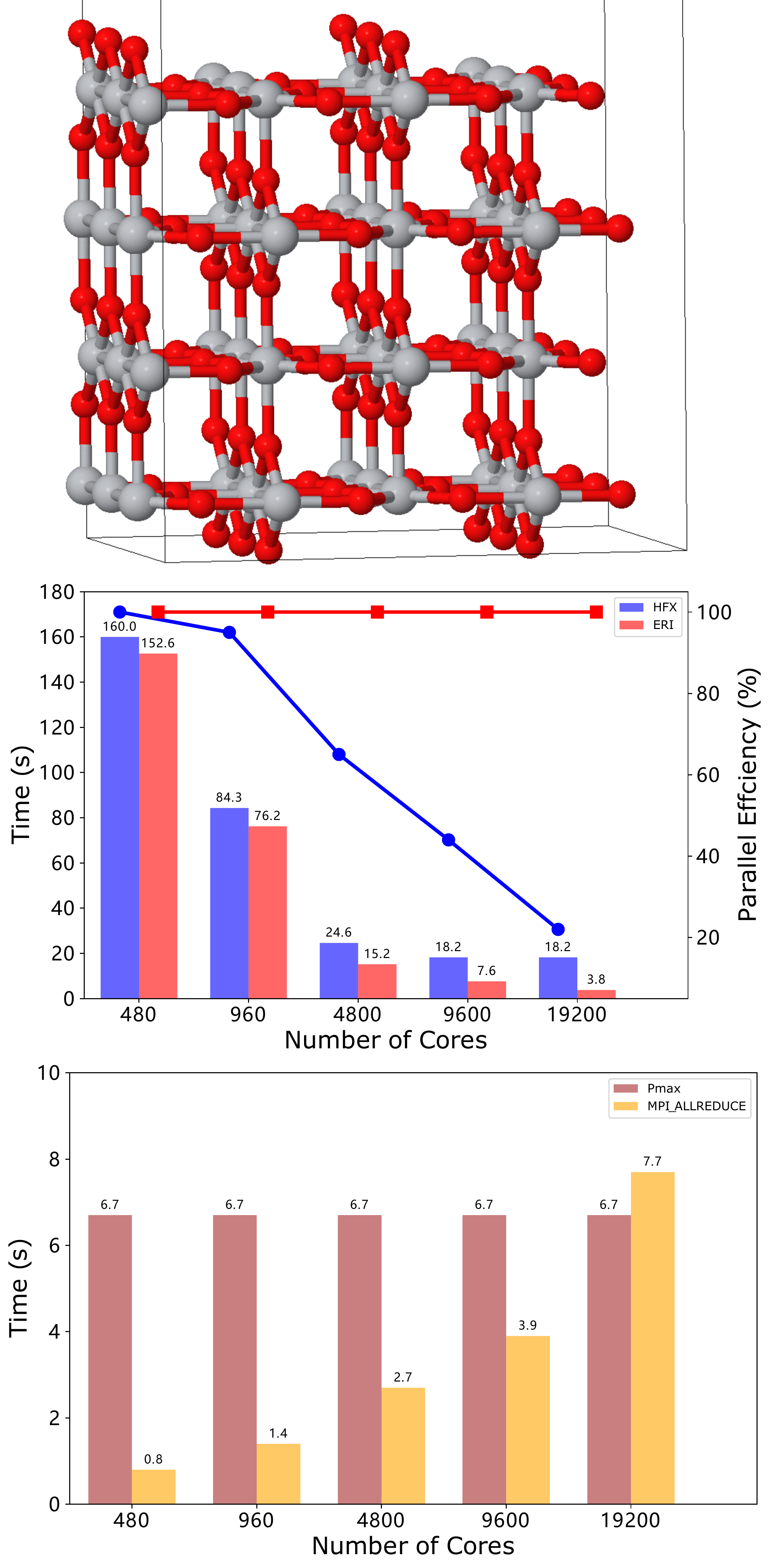}
   \caption{The strong scalability for the periodic TiO$_2$ surface system.  The blue/red/brown/orange bars correspond to the
    simulation time for the total HFX construction/ERI calculation/$P_{max}$ construction/MPI\_ALLREDUCE. The parallel efficiency of the HFX construction is labels with blue circles which the parallel efficiency of the ERIs calculations is labels with red squares. The time are annotated on top of the bars. The difference between the HFX time and the ERI time comes from the contributions from the  $P_{max}$ selection process and the {\rm MPI\_ALLREDUCE} operation for the HFX matrix which are shown in the lower panel. The discussion about the
    low 20\% HFX parallel efficiency at 19200 cores for this example is given in the text.}  
  \label{fig:TiO2}
 \end{figure}

\begin{figure}
  \includegraphics[width=0.5\textwidth]{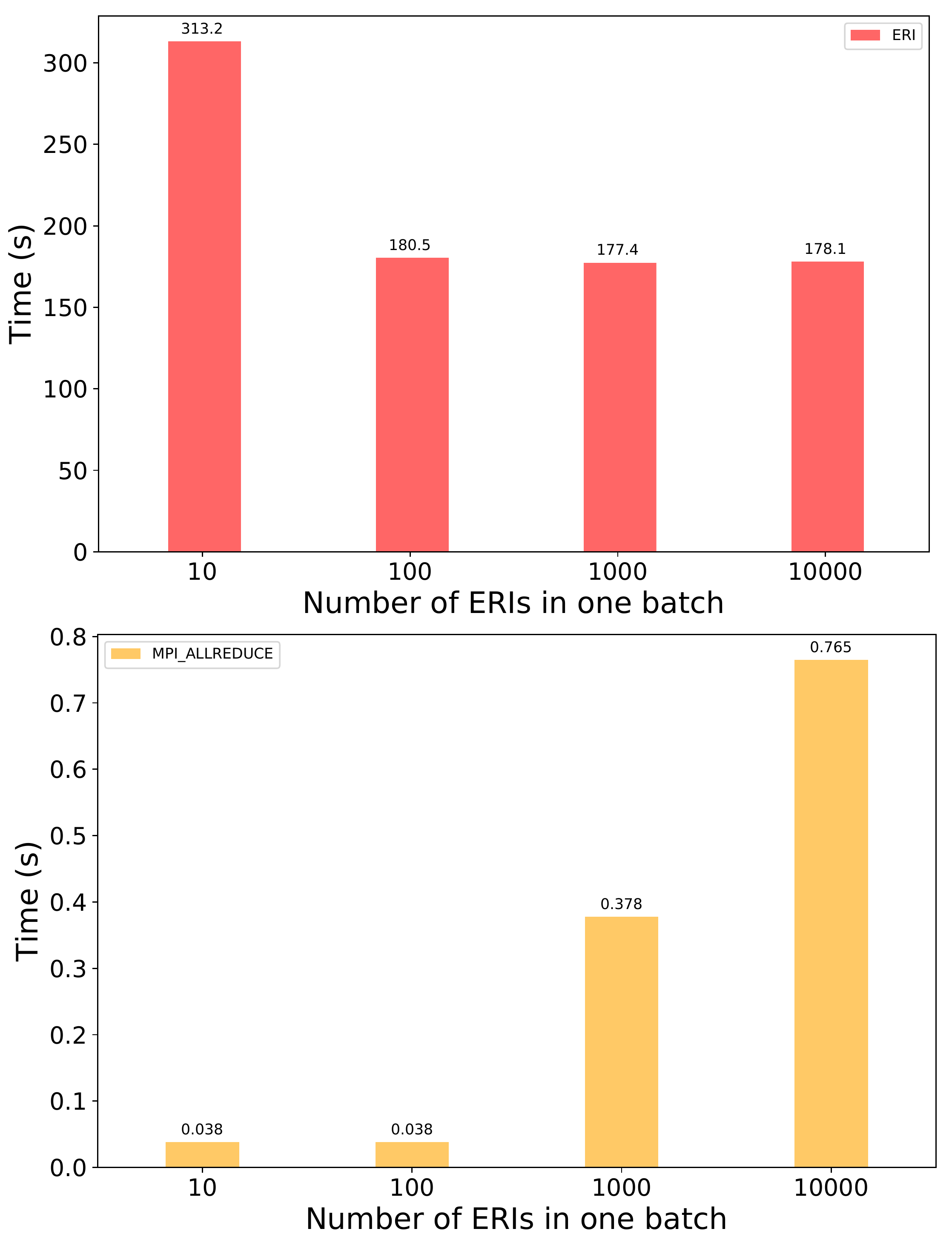}
   \caption{The time used for the periodic TiO$_2$ surface system with respect to the number of the
   ERIs in one batch. Here 480 cores are used. The red bars correspond to the
    simulation time for the ERI calculation, which are shown in the upper panel; the orange bars 
    correspond to the {\rm MPI\_ALLREDUCE}, which are shown in the lower panel. The time are annotated on top of the bars.  }  
  \label{fig:TiO2-nblock}
 \end{figure}

\begin{figure}
  \includegraphics[width=0.5\textwidth]{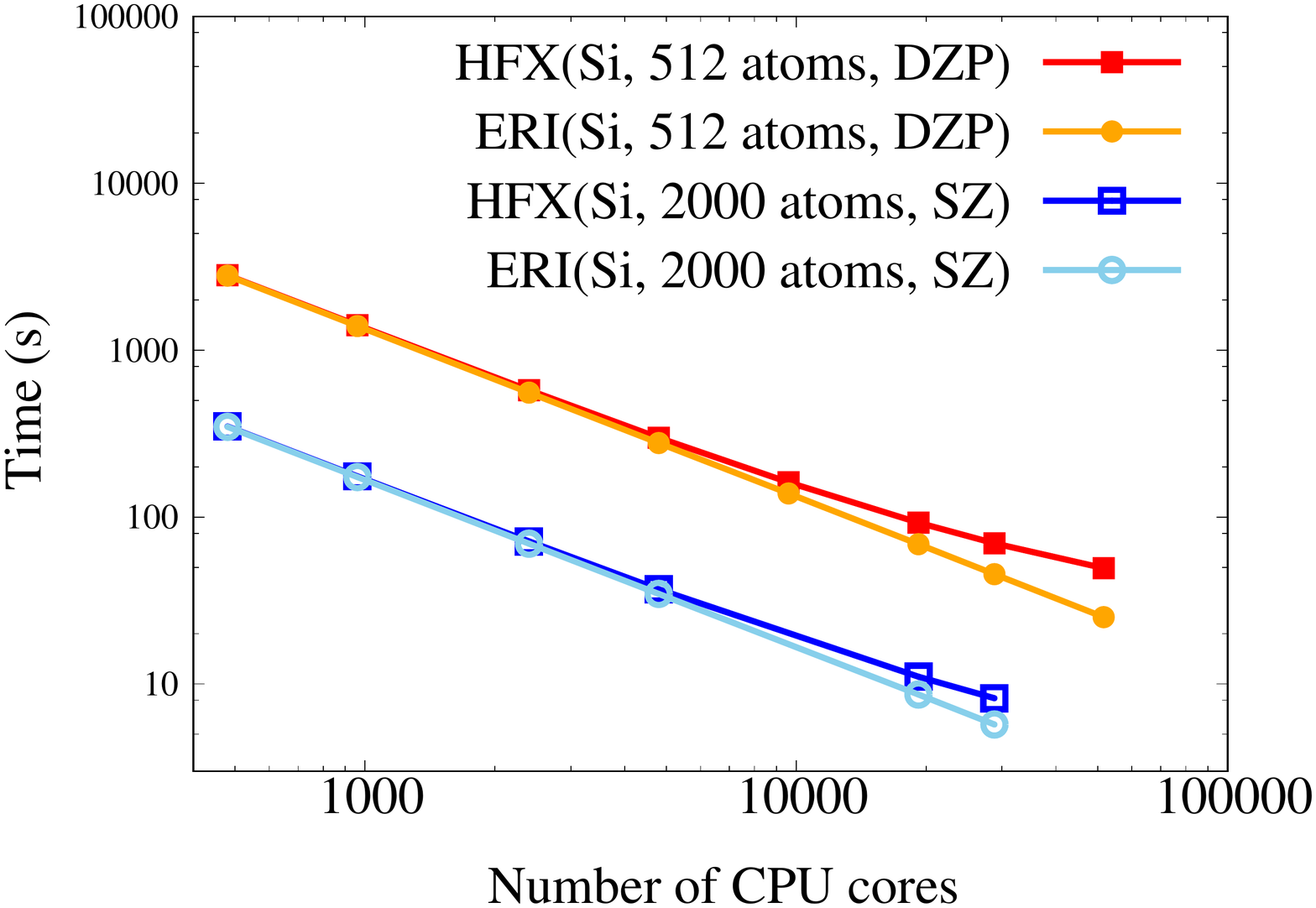}
  \caption{The strong scalability for the periodic silicon solid system with different system sizes and the basis set. }
\label{fig:Si}  
\end{figure}

The second example is a TiO$_2$ surface system supercell consisting of 144 atoms. The Ti and O atoms are described using single-$\zeta$ plus polarization (SZP) valence basis sets yielding 1488 atomic orbitals within the unitcell and 13392 atomic orbitals within the supercell. It should be noted
that the SZP calculations underestimate the electronic bandgap by roughly 8\% with respect to the DZP basis set for the TiO$_2$ bulk. Here we use SZP for the TiO$_2$ surface system just to evaluate the parallel efficiency since the scalability does not depend on the basis set as shown in Fig.\ref{fig:Si}. The calculations have been performed using 6$\times$6$\times$10 k-points in the primitive Brillouin zone. In Fig.~\ref{fig:TiO2}, the total runtime for the HFX construction in one SCF cycle and the contributions from the calculation of the two-electron integrals (i.e. ERI), the calculation of the maximal values of density matrix  in 
each shell at every SCF cycle(i.e. $P_{max}$) and the global summation of the HFX matrix(i.e. MPI\_ALLREDUCE) are displayed. Comparing to the 1 dimension DNA system,  the parallel scaling of the
 calculation of the ERIs is again near ideal with 100\% parallel efficiency,  which takes almost 95\% time with 480 cores, on the other hand, the time for $P_{max}$ and  MPI\_ALLREDUCE do not scale with the number of CPU cores, so as the number of the CPU cores increased, 
the  $P_{max}$ time percentage change from 4\% to 37\%, while the MPI\_ALLREDUCE time percentage change from 1\% to 42\%. As a result,
although the parallel efficiency of ERIs at 192000 cores is nearly 100\%, the parallel efficiency of the total HFX time which 
including ERI,  $P_{max}$, and MPI\_ALLREDUCE is only 20\% at 192000 cores. Here the very low 20\% parallel efficiency comes from two reasons: firstly, the calculation of   $P_{max}$ is not distributed, which accounts for 37\% of the total time at 192000 cores; secondly, the MPI\_ALLREDUCE communication time is relatively long, accounting for 42\% of the total time at 19200 cores. It should be noted that, the MPI\_ALLREDUCE communication time can be dramatically reduced by changing the number of ERIs in one
batch,which we called $\rm{n\_block}$ in the flowing, as shown in Fig.\ref{fig:TiO2-nblock}. When decreasing the $\rm{n\_block}$ value from
10000 to 100, the communication time can be reduced by 20 times, this is because the $\rm{n\_block}$ could affect the balance of the number of the ERIs in each core, and finally  the communication time.  If we use an optimal value of the $\rm{n\_block}$ and also distribute the calculation of $P_{max}$ , then the parallel efficiency should be increased.

\begin{figure} 
  \includegraphics[width=0.5\textwidth]{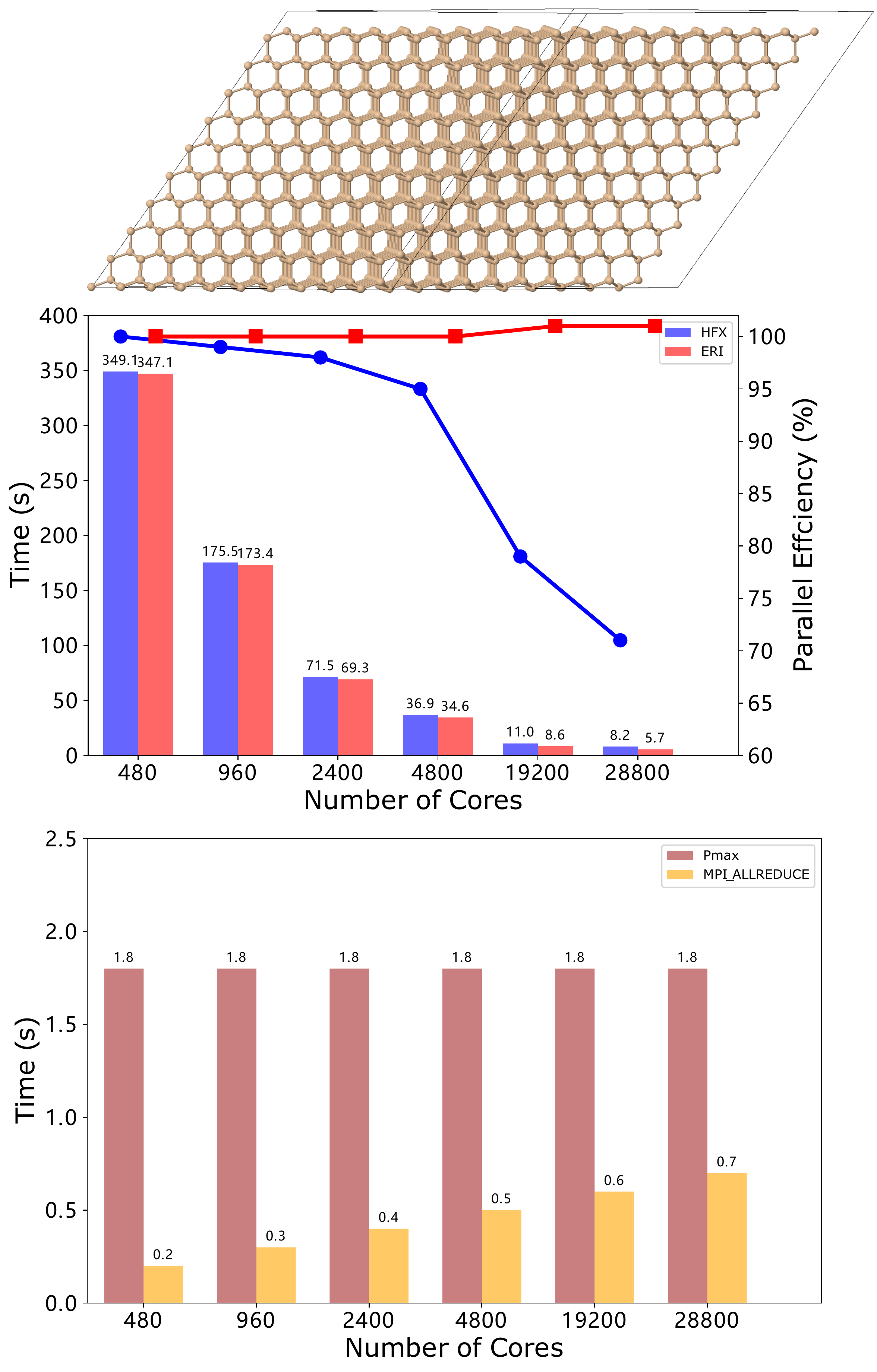}
  \caption{The stronge scalability for the periodic silicon solid system.  The blue/red/brown/orange bars correspond to the
    simulation time for the total HFX construction/ERI calculation/$P_{max}$ construction/MPI\_ALLREDUCE. The parallel efficiency of the HFX construction is labels with blue circles which the parallel efficiency of the ERIs calculations is labels with red squares. The time are annotated on top of the bars. The difference between the HFX time and the ERI time comes from the contributions from the  $P_{max}$ selection process and the {\rm MPI\_ALLREDUCE} operation for the HFX matrix which are shown in the lower panel.  }  
  \label{fig:Si-sz}
\end{figure}

The third example is a silicon (10$\times$10$\times$10) supercell consisting of 2000 atoms, a single-$\zeta$ basis set for Si atom is adopted. One k-point is used to sample the reciprocal space due to the large unit cell.  In Fig.~\ref{fig:Si-sz}, the runtime for the HFX construction in one SCF cycle and the contributions from the calculation of the two-electron integrals (i.e. ERI), the calculation of the maximal values of density matrix in 
each shell at every SCF cycle(i.e.  $P_{max}$) and the global summation of the HFX matrix(i.e. MPI\_ALLREDUCE) are displayed. Comparing this system to the second example, the integral calculation maintains perfect scalability over the whole range of CPU cores that considered from 480 to 28800, while now the parallel efficiency of the HFX  matrix construction step is dramatically improved to 80\% at 19200 cores,  this is because in this case, the  $P_{max}$ and MPI\_ALLREDUCE time percentage is 16\% and 5\% respectably.  The improved scalability in this example is basically owning to the more balancing distribution of the number of the ERIs as only the s-type and p-type orbitals have been considered. On the other hand, in the first and the second
examples, the polarization d-type orbitals have been involved, which caused the imbalance of the number of the ERIs as well as the corresponding HFX matrix elements, and increase their time of the communication costs~(MPI\_ALLREDUCE).
Finally, in this third example, the parallel efficiency of ERIs at 28800 cores is nearly 100\%, while the parallel efficiency of the total HFX time is 70\% at 28800 cores.  

In order to see the influence of the basis set, the parallel scalability for a unit cell containing 2000 Si atoms with SZ basis set and 512 Si atoms with DZP basis set is shown in Fig.~\ref{fig:Si}. All calculations again use one k-point to sample the reciprocal space due to the large unit cell.  Here we can see for both the SZ and DZP basis set, almost ideal scaling is achieved for the ERI calculations,  however, at 28800 cores, the communication costs~(MPI\_ALLREDUCE) for SZ/DZP basis set are 0.7s/12.6s respectively, which
caused the final parallel efficiency of the total HFX time is 67\%/70\% at 28800 cores. The largest number of CPU cores we have tested is 51600 for the Si consisting of the 512 atoms with DZP basis set, and in this case, duo to the communication time increased to 12.8s with the percentage 24\% of the total HFX time, the parallel efficiency of the HFX is 53\% although the integral calculation maintains perfect 100\% scalability. 
 
\section{Conclusion}
\label{sec:conclusion}
In summary, we have shown our dynamic parallel algorithms for the ERIs calculations based on the real-space NAO2GTO framework. We have also analyzed the performance of the parallel algorithms for parallel efficiency. Based on our results, 
the dynamic distribution of ERI shell quartet can yield very high
load balance, nearly ideal 100\% parallel efficiency for the calculation of ERIs has been achieved. However, because the  $P_{max}$ selection and the summation
of the HFX matrix procedures do not distribute over CPU cores, the parallel efficiency of the total HFX construction is not as good as the calculation of the ERIs. On the next step, we need to also distribute the $P_{max}$ calculation by using the dynamic parallel distribution algorithm to improve the parallel efficiency. Furthermore, a shared memory method with one-sided commutation method should also be adopted to improve the performance. Such a two-level master-worker dynamic parallel distribution algorithm proposed in this work can also be extended to adopt the graphics processing units (GPUs) as accelerators.

\bibliography{mybibfile}

\section{Acknowledgments}
This work is supported by the Special Fund for Strategic Pilot Technology of Chinese Academy of Sciences (XDC01040000). The authors thank the Tianhe-2 Supercomputer Center for computational resources.
\end{document}